\documentclass[preprint,authoryear,12pt]{elsarticle}

\usepackage{graphics}
\usepackage{url}

\usepackage{amssymb}
\usepackage{natbib}
\usepackage[ps2pdf,%
a4paper=true,%
breaklinks=true,%
colorlinks=true,%
pdfauthor={First Author et al.},%
pdftitle={Template for manuscripts in Advances in Space Research}%
]{hyperref}

\journal{Advances in Space Research}

\makeatletter
\def\@author#1{\g@addto@macro\elsauthors{\normalsize%
    \def\baselinestretch{1}%
    \upshape\authorsep#1\unskip\textsuperscript{%
      \ifx\@fnmark\@empty\else\unskip\sep\@fnmark\let\sep=,\fi
      \ifx\@corref\@empty\else\unskip\sep\@corref\let\sep=,\fi
      }%
    \def\authorsep{\unskip,\space}%
    \global\let\@fnmark\@empty
    \global\let\@corref\@empty  
    \global\let\sep\@empty}%
    \@eadauthor={#1}
}
\makeatother

\begin{document}

\begin{frontmatter}



\title{Recent advancements in the EST project}


\author{Jan Jur\v{c}\'{a}k\corref{cor}}
\address{Astronomical Institute of the Academy of Sciences, Fri\v{c}ova  298, 25165 Ond\v{r}ejov, Czech Republic}
\cortext[cor]{Corresponding author}
\ead{jurcak@asu.cas.cz}


\author{Manuel Collados}
\address{Instituto de Astrof\'isica de Canarias (IAC), V\'ia Lact\'ea, 38200 La Laguna (Tenerife), Spain\\
Departamento de Astrof\'isica, Universidad de La Laguna, 38205 La Laguna (Tenerife), Spain}

\author{Jorrit Leenaarts}
\address{Institute for Solar Physics, Department of Astronomy, Stockholm University, AlbaNova University Centre, SE-106 91 Stockholm, Sweden}

\author{Michiel van Noort}
\address{Max-Planck Institute for Solar System Research, Justus-von-Liebig-Weg 3, 37077 G\"ottingen, Germany}

\author{Rolf Schlichenmaier}
\address{Kiepenheuer-Institut f\"{u}r Sonnenphysik, Sch\"{o}neckstr. 6, 79104 Freiburg, Germany}

\begin{abstract}

The European Solar Telescope (EST) is a project of a new-generation solar telescope. It has a large aperture of 4~m, which is necessary for achieving high spatial and temporal resolution. The high polarimetric sensitivity of the EST will allow to measure the magnetic field in the solar atmosphere with unprecedented precision. Here, we summarise the recent advancements in the realisation of the EST project regarding the hardware development and the refinement of the science requirements.

\end{abstract}

\begin{keyword}
instrumentation \sep the Sun \sep EST
\end{keyword}

\end{frontmatter}

\parindent=0.5 cm

\section{Overview of the EST project development}


The development of EST is organized by the European Association for Solar Telescopes (EAST). The EAST was formed in June 2006 and is currently a consortium of 23 solar research centres in 17 European countries. To ensure access of European solar physicists to world-class observing facilities, the primary goal of the EAST consortium is to develop, construct and operate the next generation 4-metre class solar telescope on the Canary Islands.

The conceptual design study of EST was realised in the frame of the FP7 project ``EST: The large aperture European Solar Telescope''\footnote{http://istar.ll.iac.es/files/58cef4ec1579d9e39754c76d5.pdf} supported by the European Commission. This project was realised between February 2008 and July 2011 and involved 14 research institutions and 15 industrial partners. The conceptual design phase was summarised in the ``EST: Conceptual Design Study Report'' containing the proposed solutions for the telescope itself and all its subsystems along with management plans, socio-economic impact, and financial feasibility. A detailed ``Report on technical, financial, and socio-economic aspects''\footnote{www.est-east.eu/est/images/media/pdf/EST\_socio-economic\_web\_version.pdf} was also one of the conceptual phase project deliverables. The conceptual design of the EST has been positively evaluated in 2011 by a panel formed by prestigious external reviewers.  

The follow up project of the European solar physics community to prepare for the EST research infrastructure (RI) was the SOLARNET (High-Resolution Solar Physics Network), a network aimed at bringing together and integrating the major European research infrastructures in the field of high-resolution solar physics in order to promote their coordinated use and to secure the future development of the next-generation RI in the form of EST. The project was carried out between April 2013 and March 2017 and funded from the EC Integrated Infrastructures Initiative. It involved 32 partners from 16 countries: 24 EU research institutions, 6 EU private companies, and 2 USA research institutions.

GREST (Getting Ready for EST) is an on-going project, funded by the EC H2020 program from June 2015 till June 2018. GREST is taking the EST to the next level of development by undertaking crucial activities to improve the performance of current state-of-the-art instrumentation. Also legal, industrial, and socio-economic issues are addressed in the frame of this project. GREST involves 13 partners from 6 EU countries, 3 of them are private companies.

EST formally entered the active project list of the ESFRI (the European Strategy Forum on Research Infrastructures) roadmap in March 2016 as the flagship project for the European Solar Physics community. Acceptance of EST project to the list of ESFRI roadmap is an important milestone in the implementation of the EST as the primary task of ESFRI is to support and help the projects on the roadmap move towards realisation.

On 1st April 2017, the PRE-EST project (Preparatory phase for EST) was commenced. It aims to provide both the EST international consortium and the funding agencies with a detailed plan regarding the implementation of EST. Moreover, PRE-EST will lead the detailed design of the EST key elements to the required level of definition and validation for their final implementation. This project is partly funded under the H2020 framework till March 2021. One of the major decisions in this ongoing project is the plan to establish an ERIC (European Research Infrastructure Consortium) as the legal entity of the EST RI. The plan is to establish the ERIC with Spain as the host state.

Currently, there is a proposal for continuation of the SOLARNET project applying for the call H2020-INFRAIA-2018-2020. The goals of this project are comparable to the original SOLARNET project and aim at integrating the major European infrastructures in the field of high-resolution solar physics. 

The most recent information about the project development and status can be found on the project web-site\footnote{\url{http://www.est-east.eu}}, followed on various social media like Facebook\footnote{\url{https://www.facebook.com/EuropeanSolarTelescope}}, Twitter\footnote{\url{https://twitter.com/estsolarnet}}, Youtube\footnote{\url{https://www.youtube.com/user/ESTtvCHANNEL}}, and Linkedin\footnote{\url{https://www.linkedin.com/company/european-solar-telescope/}}.

\section{Overview of the EST science goals}

The EST's main research goal is to observe the Sun. The Sun is the only star that can be studied in high resolution. We can study fundamental interactions between plasma, magnetic field, and radiation in the solar atmosphere. Although we are not able to spatially resolve the intrinsic scales of these magneto-hydrodynamic processes yet, the magnetic Sun forms the basis of our knowledge of the cosmic magnetic field. With the EST we will improve the resolution and enhance our understanding of fundamental science.

A variety of structures and physical processes can be observed on the Sun. These phenomena are connected by a variable magnetic field, which becomes most prominent during the maximum of the Sun's magnetic activity: Energetic events are the result of interaction of the magnetic field with ionised plasma and its radiation field. The EST will focus on observing this interaction. Magnetic fields are detected and characterised by analysing their imprint on the polarisation, therefore EST is designed to achieve the polarimetric accuracy of $10^{-4}$.

The optical design and instruments of the EST are optimised for observing the physical coupling of the atmospheric layers from the deepest photosphere to the upper chromosphere. Suitable techniques have been developed to determine the thermal, dynamic, and magnetic properties of the plasma over many scale heights with imaging, spectroscopy, and spectro-polarimetry instrumentation. The EST will be equipped with a suite of instruments to simultaneously observe in various wavelengths from 380~nm to 2~$\mu$m, so that the solar photon flux can be exploited more efficiently than at other current ground-based or space telescopes.

\section{Hardware development}
\label{hardware}

The design of EST and its sub-systems was summarised in the conceptual design study. However, the detailed design of the sub-systems is now in progress and feasibility of the individual sub-systems is studied. There are studies on the purely technical sub-systems like the heat rejecter at the primary focus of the EST \citep{Berrilli:2010, Berrilli:2017}. Studies on the sub-systems that directly influence the observations, e.g., the multi-conjugate adaptive optics that can correct for the atmospheric aberrations in the whole field of view of the telescope \citep{Stangalini:2010, Soltau:2010, Montilla:2012, Montilla:2013, Montilla:2014, Rosa:2016, Stangalini:2016}. 

\begin{figure}
\includegraphics*[width=\linewidth]{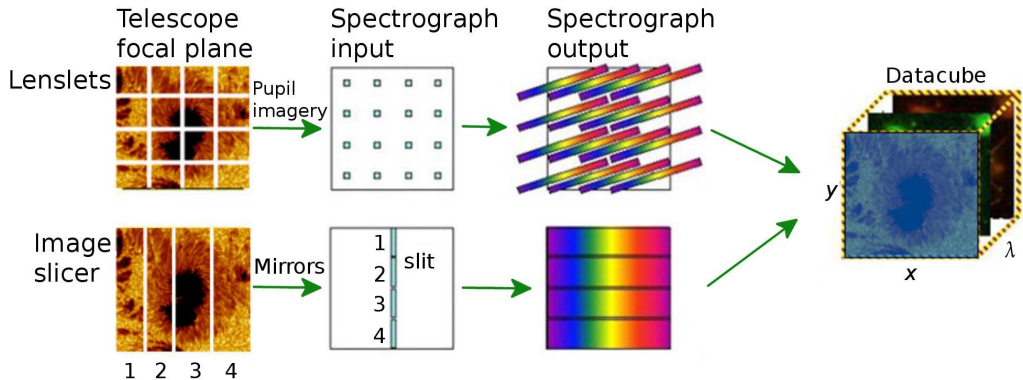}
\caption{Schemes of the IFU units that are developed for EST. Adapted from \citet{Calcines:2013}.}
\label{ifu_scheme}
\end{figure}

We describe here in more details the development of instruments that will allow for integral field spectropolarimetry, i.e., simultaneous observations in both bi-dimensional spatial and spectral domain. Schemes of two technical solutions of these integral field units (IFU) that are under development for the EST are shown in Fig.~\ref{ifu_scheme}.

\subsection{Micro-lens array}

One of the options is so-called micro-lensing, which is developed at MPS, primarily by M. van Noort. The principle of this technology is shown in the upper row of Fig.~\ref{ifu_scheme}. Every pixel (resolution element) at the telescope focal plane is re-imaged to a smaller area, the resulting sparse image is fed to a spectrograph, and the dispersion angle of the spectrograph is then optimised to minimise the mixture of spatial and spectral information. This concept necessitates the use of a narrow-band pre-filter to ensure that the spectra of different spatial pixels do not overlap. Since the micro-lens array is a refraction-based device, it is not achromatic. The periodic design unavoidably produces Moire fringes and the instrument must be kept in stable conditions so that the fringes do not change in time. On the other hand, the optical design of micro-lens arrays is very simple and massively parallel, allowing it to be scaled to the order of $10^5$ image elements. In addition, since the image has already been sampled by the micro-lens array, the optical quality of most of the optical elements after the micro-lens array is not very critical for the image quality.

\begin{figure}
\begin{center}
\includegraphics*[height=4.3cm]{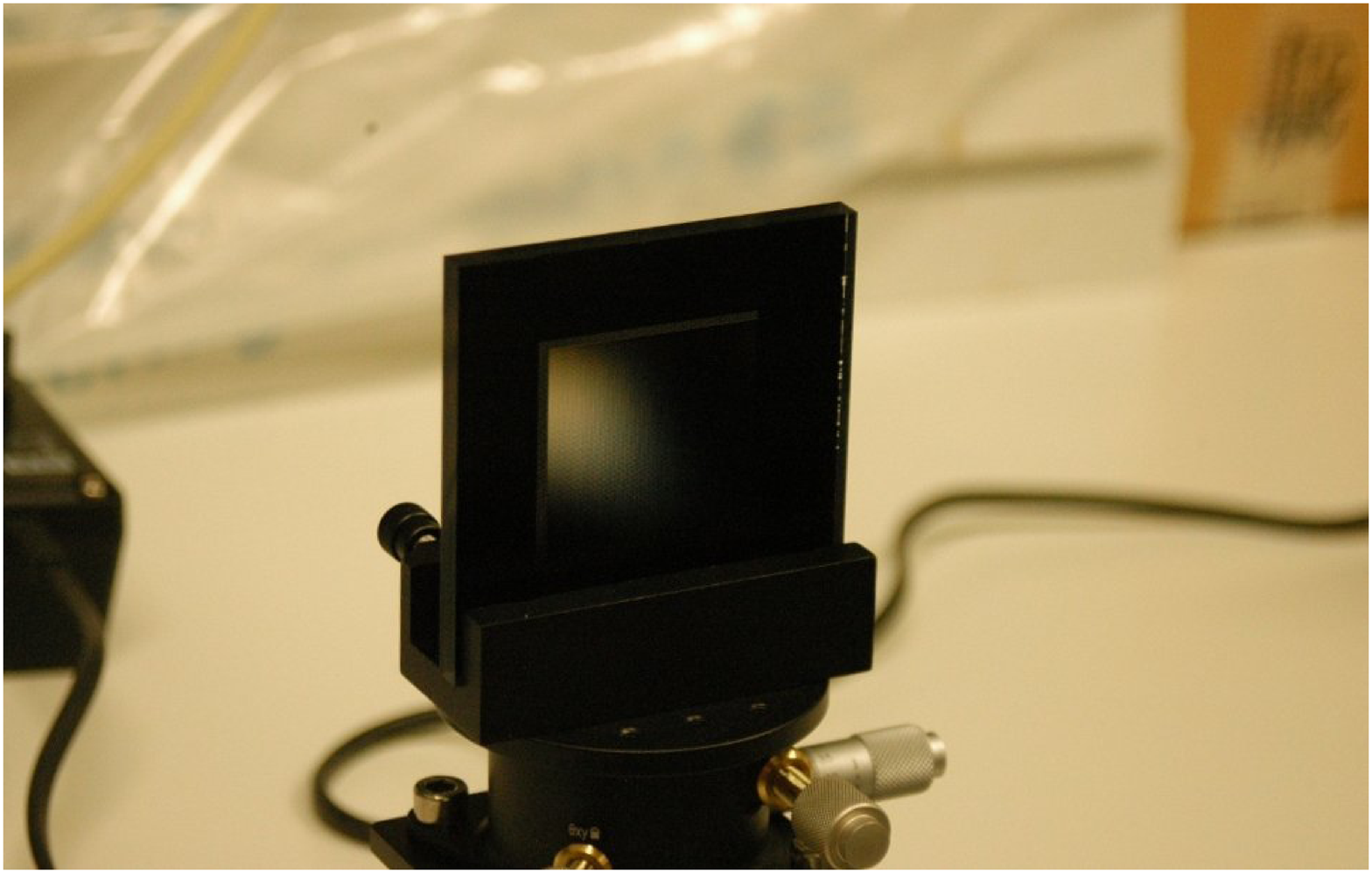}
\includegraphics*[height=4.3cm]{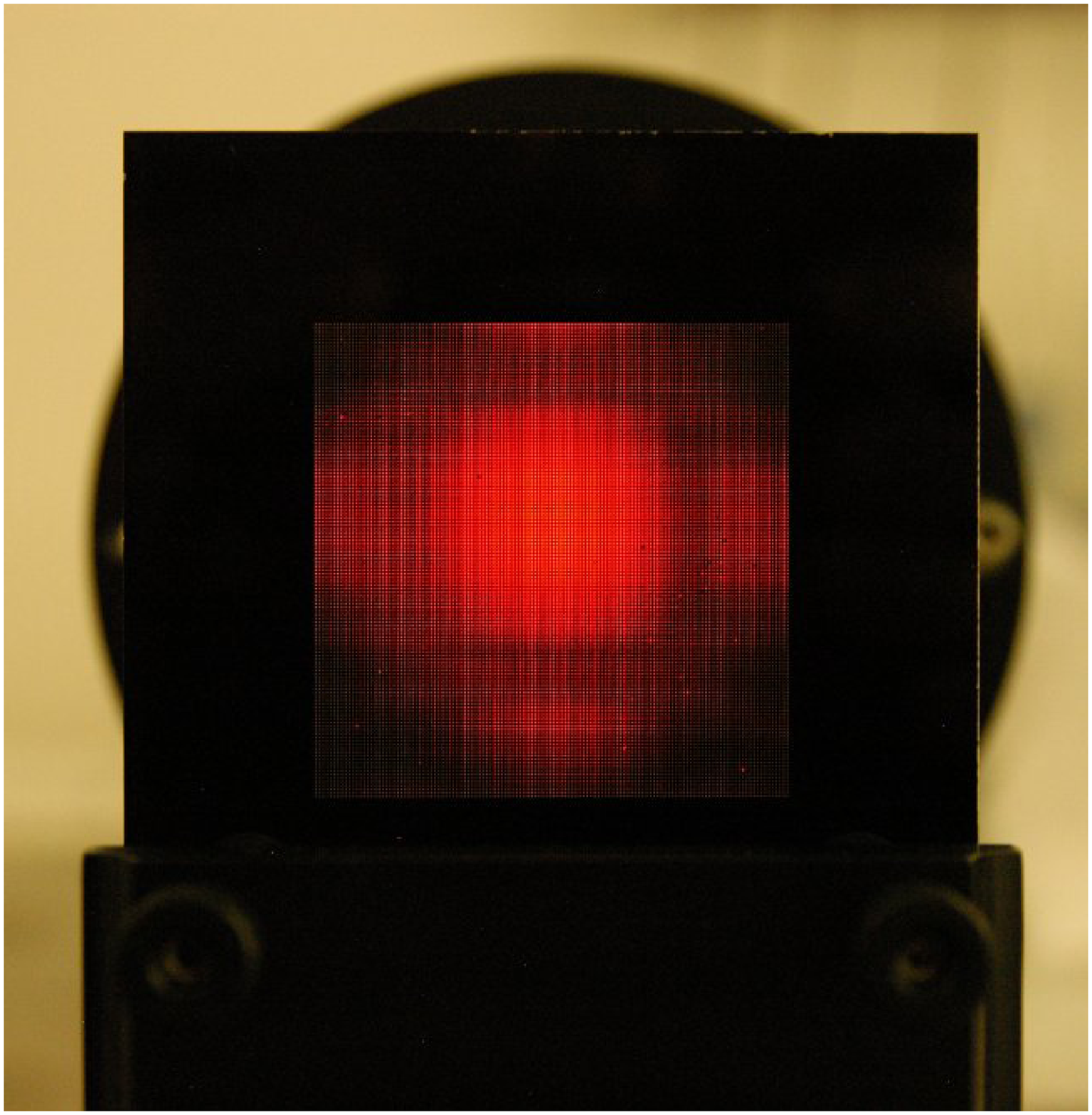}
\end{center}
\caption{Photos of the prototype of the micro-lens array tested at SST by M. van Noort.}
\label{micro-lens-array}
\end{figure}

\begin{figure}
\includegraphics*[width=\linewidth]{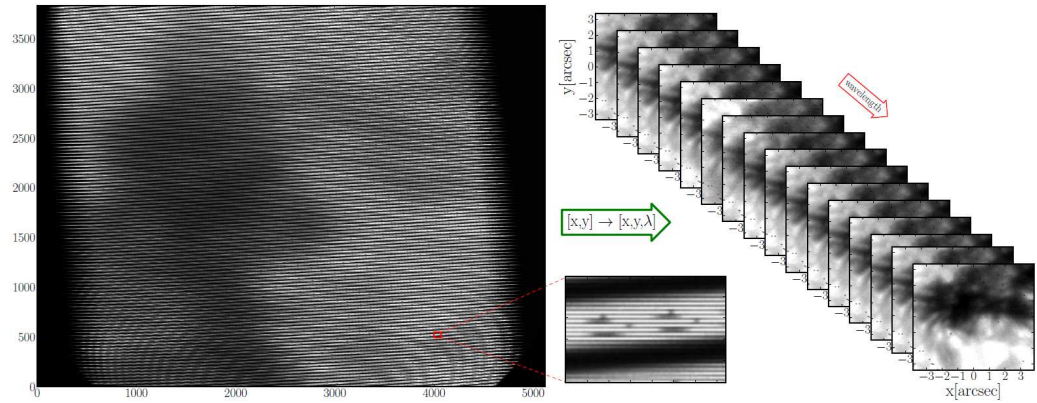}
\caption{Test data obtained with the micro-lens array at SST. Image of the whole CCD chip is on the left, amplified sample in the middle, and reconstructed images of the field of view on the right.}
\label{MLA_data}
\end{figure}

The concept of micro-lens array was already tested by Y. Suematsu \citep{Suematsu:1999, Suematsu:2011} using a single micro-lens array, who found his instrument to have very limited performance. This was primarily due to his specific optical configuration, which caused the image elements to be re-imaged onto a grating, resulting in a variable, generally poor, spectral performance. To avoid this problem, a second micro-lens array is needed and it must be aligned to the first one with formidable accuracy. The two micro-lens arrays are therefore fabricated on two sides of a single substrate to ensure accurate alignment (see Fig.~\ref{micro-lens-array}), and it in addition contains a straylight mask. Manufacturing of such a device has become technologically possible only recently and a prototype of this instrument was constructed by M. van Noort and tested on the Swedish Solar Telescope (SST). 

Some results of the test observations are shown in Fig.~\ref{MLA_data}, where the image on the left corresponds to the whole sensor image, and a sample of the spectra at individual spatial pixels are shown in the middle. These data are then rearranged to create spatial maps at individual wavelengths as shown in the right part of Fig.~\ref{MLA_data}. Details about the instrument will be summarised in an up-coming paper from M. van Noort.

\subsection{Image slicer}

The other option that is considered for an IFU for EST is a so-called image slicer. Its concept is shown in the bottom row of Fig.~\ref{ifu_scheme}. An image slicer uses an array of thin mirrors at the focal plane to redistribute the light from the entrance field of view along one long spectrograph slit. Since the image slicer has a mirror-based design, it is in principle achromatic. Each slit retains an image information, which is not periodic, therefore there are no Moire fringes. If single-slit version of the IFU is used, there is no overlap of the spectra. However, this advantage is lost once the thin mirrors redistribute the light to number of slits. This IFU design is very challenging optically. Each slit requires slit-specific optics and alignment. The optical quality has to be high throughout the instrument as image information is propagated. This IFU is developed by IAC and details about this instrument can be found in \citet{Calcines:2013, 2014SPIE.9147E..3IC}. 

\begin{figure}
\begin{center}
\includegraphics*[height=3.3cm]{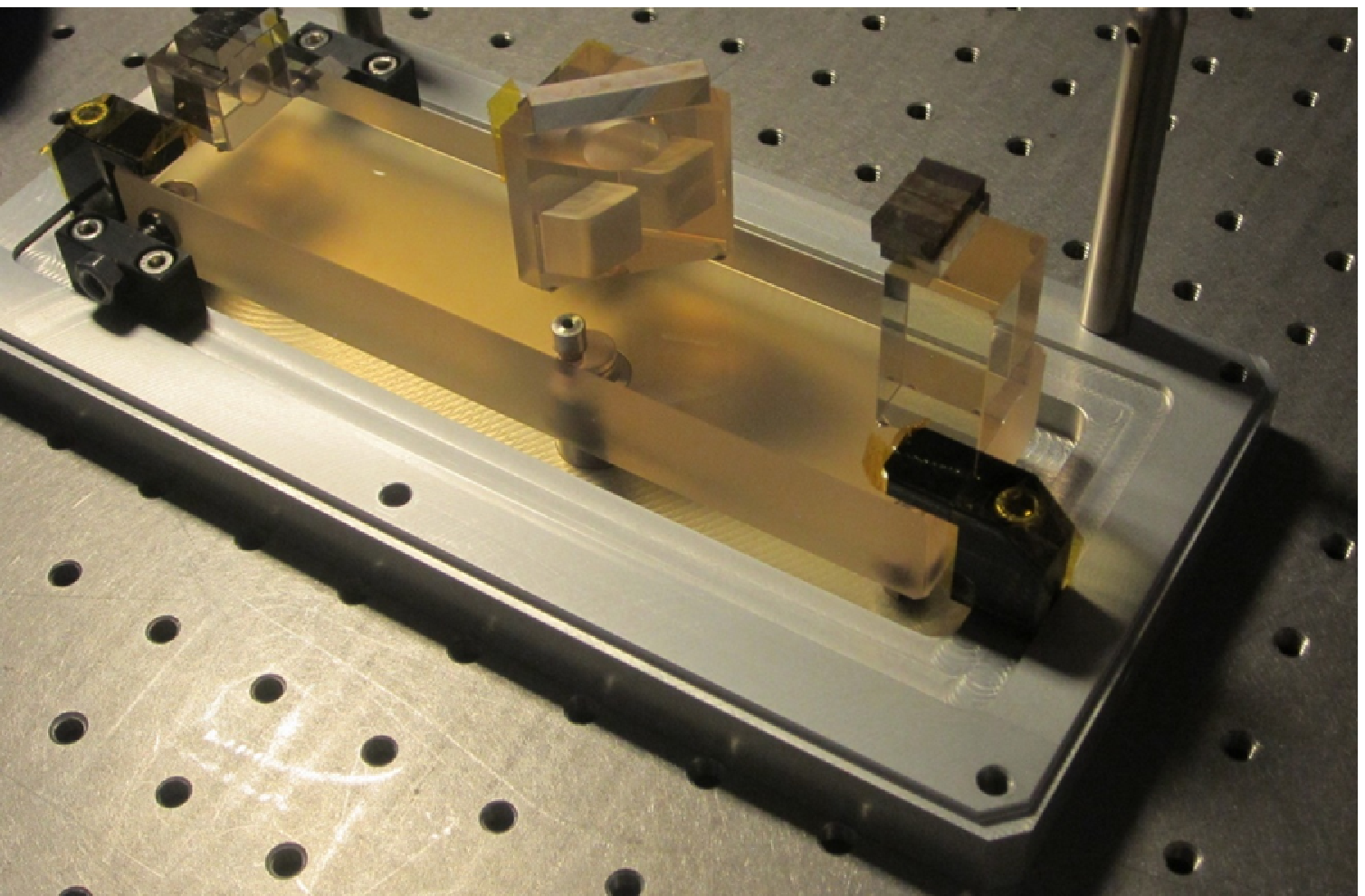}
\includegraphics*[height=3.3cm]{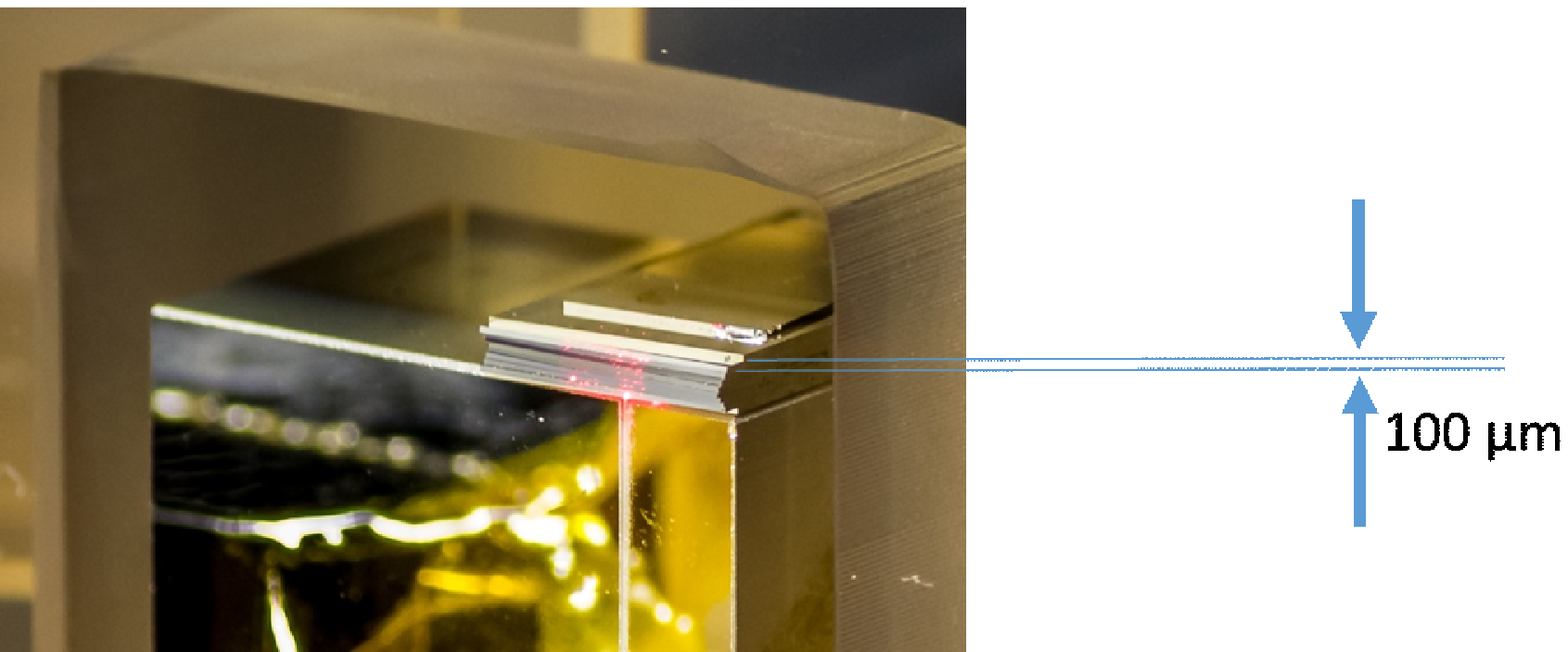}
\end{center}
\caption{Photos of the prototype of the image slicer developed at IAC and tested on GREGOR.}
\label{image_slicer_photo}
\end{figure}

\begin{figure}
\begin{center}
\includegraphics*[height=4cm]{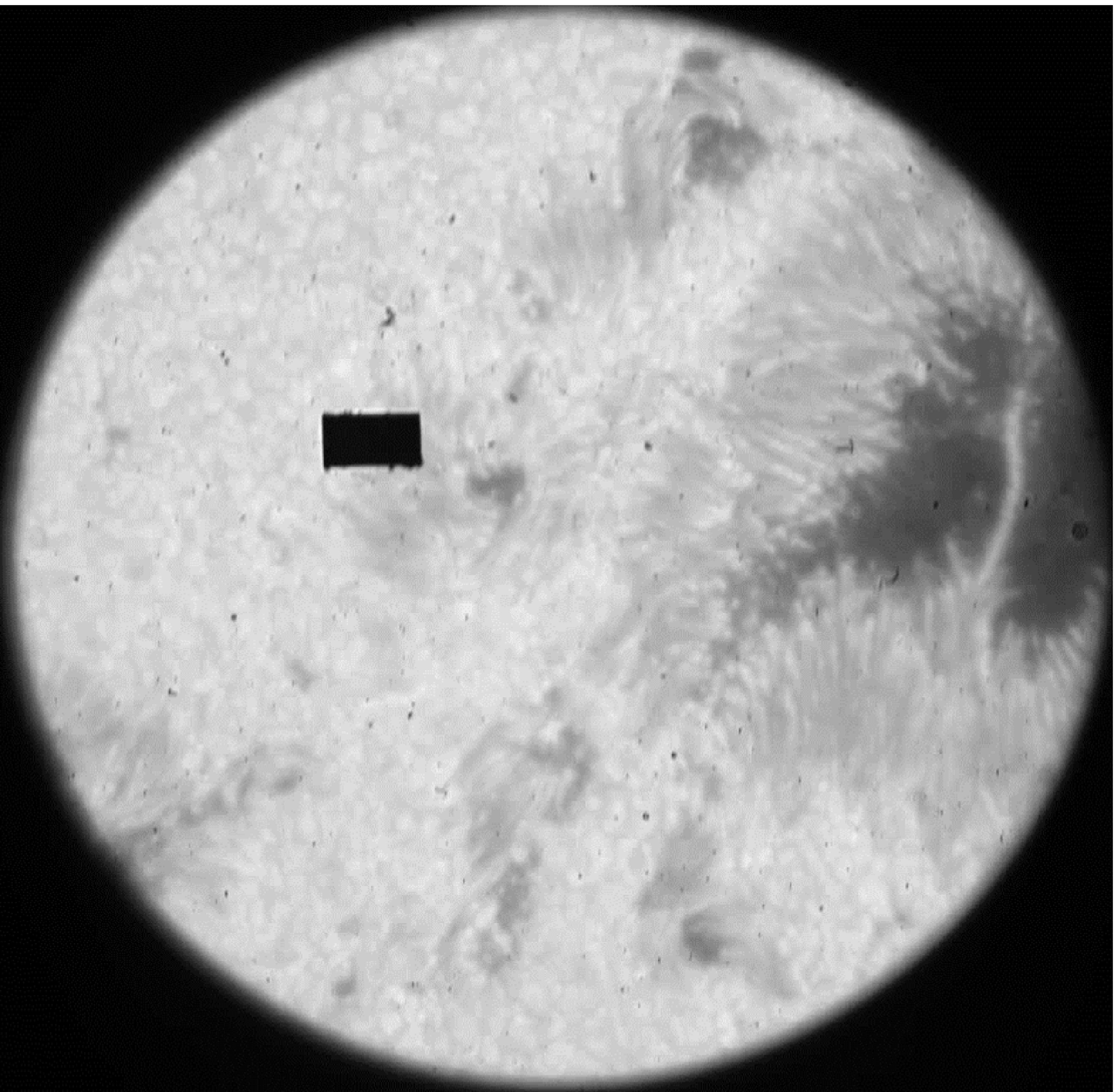}
\includegraphics*[height=4cm]{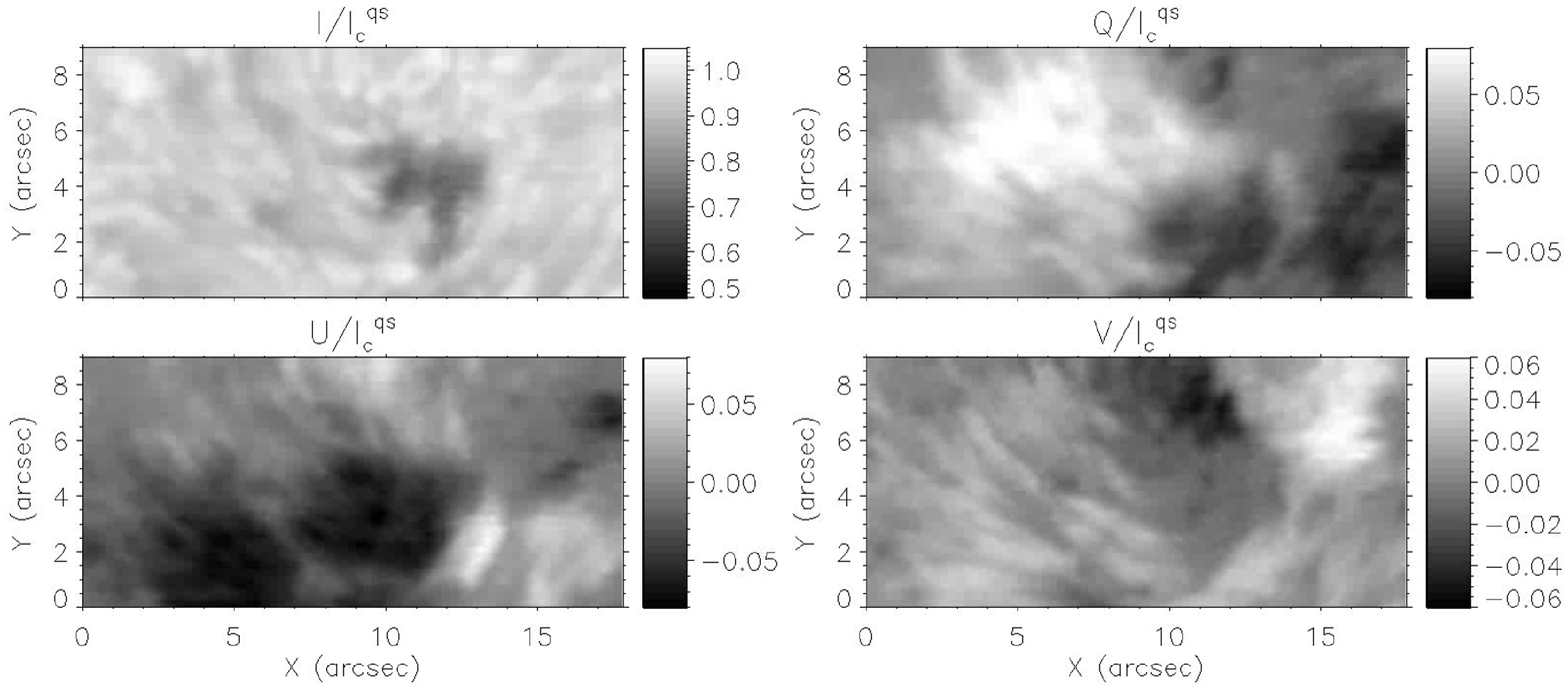}
\end{center}
\caption{Test data obtained with the image slicer IFU at GREGOR. ''Slit-jaw`` image on the left shows the size of one tile. Reconstructed maps of $I$, $Q$, $U$, and $V$ intensities from a FOV composed of $3 \times 3$ IFU tiles are shown on the right.}
\label{image_slicer_data}
\end{figure}

In Fig.~\ref{image_slicer_photo} we show the prototype of the image slicer for the EST that was tested at the GREGOR telescope. In Fig.~\ref{image_slicer_data} there are snapshots from the test observations made with the prototype of image slicer IFU attached to the GREGOR telescope. On the left is the image from the "slit-jaw`` camera depicting the field of view of one IFU tile. Panels on the right are maps of $I$, $Q$, $U$, and $V$ intensities from a field of view composed of $3 \times 3$ IFU tiles. Note that the image slicer provides simultaneous spatial and spectral observations for a single tile and it took nine sequential measurements to construct the right panels in Fig.~\ref{image_slicer_data}.

\section{Update of the Science Requirement Document}

The design of the EST including the set-up of the light distribution system and the set of instruments intended for the first light is summarised in the ``EST: Conceptual Design Study Report'' and was based on the Science Requirement Document (SRD) that was also developed during the conceptual design study of EST. 

The advancements in the hardware development of new instruments (like those described in Sect.~\ref{hardware}) along with the advancements in our understanding of processes occurring in the solar atmosphere require an update of the original SRD. This update is now in progress and will be presented to the community shortly before the EST science meeting on June 11-15, 2018. This update will result into the refinement of the light distribution system and the set of post-focal instruments required for the first-light science.

However, this update will not affect the main strengths of the EST baseline design. The entire telescope is designed to be polarimetrically compensated, i.e., the M\"uller matrix of the telescope will be unity and independent on the wavelength and time. Due to the alt-azimuthal mount the field of view is rotating in time. To compensate for this effect, seven mirrors of the transfer optics before the Coud\'e focus will be arranged on a rotating platform and thus used as an optical field-of-view de-rotator. This design allows to mount the instruments on a fixed platform, which is advantageous in terms of simplicity, instrument stability, higher number of instruments, and flexibility of the instruments upgrades. Another advantage of the EST design is the height of the tower of approximately 35~m that will reduce the ground layer effect of the local seeing conditions. 

\begin{figure}
\includegraphics*[width=\linewidth]{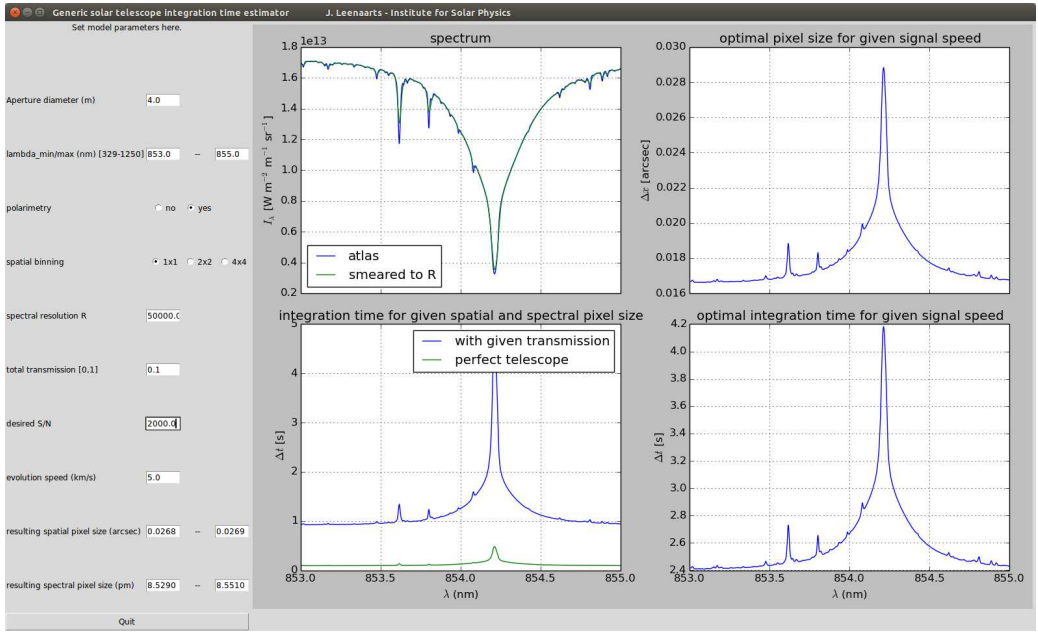}
\caption{Snapshot of the photon flux calculator developed by J. Leenaarts.}
\label{photonflux}
\end{figure}

In view of the update of the SRD, J. Leenaarts developed a photon flux calculator that provides an immense help to estimate the capabilities of the 4~m class solar telescope. It is a freely available\footnote{\url{http://dubshen.astro.su.se/\~jleen/sag/}} python routine (it is executed by photoncount.py; additional python packages might be necessary to run it). Snapshot of this photon flux calculator is shown in Fig.~\ref{photonflux}.

In the photon count calculator you can adjust:
\begin{itemize}
 \item the aperture diameter (by default set to the EST aperture of 4~m)
 \item the wavelength range of the observations
 \item polarimetric or spectroscopic mode
 \item spatial binning (by default the pixel size is set to critically sample the spatial resolution for a given wavelength)
 \item spectral resolution R ($\lambda / \Delta\lambda$)
 \item total transmission of the telescope at the given wavelength (see Table~\ref{throughput} for the throughput values of the whole system, i.e., including the transfer optics, the light distribution system, and the individual instruments. This table will be updated based on the new design of the light distribution system and the set of instruments.)
 \item desired signal-to-noise ratio
 \item evolution speed (the transversal motion of the fine structures in the solar atmosphere)
\end{itemize}

As an output, the routine provides:  
\begin{itemize}
\item resulting spatial pixel size (function of the wavelength, aperture size, and spatial binning)
 \item resulting spectral pixel size (function of the wavelength and spectral resolution)
 \item (top-left plot) the FTS atlas spectrum \citep{Neckel:1999} at the given range of observed wavelengths and the spectrum smeared to the given spectral resolution R
 \item (lower-left plot) the integration time necessary to reach the desired signal-to-noise ratio for a given spatial and spectral pixel
 \item (right plots) for these plots, the diameter of the telescope, the throughput, the spectral resolution, the desired  signal-to-noise ratio and the evolution speed are taken into account. Given these parameters, the routine computes the optimal integration time and pixel size from the constraints that a structure crosses exactly one spatial pixel during the integration time, which means that the image is not smeared owing to the evolution of solar structure.
\end{itemize}

\begin{table}
\caption{Throughput values for the current EST design.}
\begin{tabular}{llcc}
\hline
Instrument & $\lambda$ band [nm] & Individual obs. & Simultaneous obs. \\
\hline
SP	& 380-400  &   0.020	  &	0.005\\
	& 400-500  & 	   0.080 &	0.020\\	
	& 500-620 &	   0.098 &	0.025\\
	& 620-800 &	   0.098 &	0.024\\
NB1	& 380-400 &	   0.020 &	0.011	\\
	& 400-500 &	   0.080 &	0.044\\
BB1	& 380-400 & 0.019	 &	0.001\\
	& 400-500 &	   0.076 &	0.003\\
BB2	& 380-400 &	   0.017 &	0.000\\
	& 400-500 &	   0.069 &	0.001\\
BB3	& 500-620 &	   0.098 &	0.006\\
	& 620-800 &	   0.098 &	0.006\\
NB2	& 500-620 &	   0.098 &	0.046\\
NB3	& 620-800 &	   0.103 &	0.043\\
IRSP    & 800-1100  &  0.068	 &	0.014\\
	& 1100-1500   &0.088	 &	0.018\\
	& 1500-1800  & 0.090	 &	0.018\\
	& 1800-2300  & 0.090	 &	0.018\\
IRNB1	& 800-1100   & 0.061	 &	0.037\\
IRNB2	& 1500-1800  & 0.085 	 &	0.052\\
\hline
\end{tabular}
\label{throughput}
\end{table}

\section{Conclusions}
The project of the EST is advancing quickly in the recent years thanks to the joint effort of people united in the EAST consortium and those working on the realisation of projects like SOLARNET, GREST, and PRE-EST. We are entering a crucial phase of the project when the commitment of the funding agencies of individual member states have to be secured for the preparatory phase, the construction phase, and the operating costs.

\section*{Acknowledgements}
This work is carried out as a part of the project Preparatory phase of the EST (PRE-EST) funded by the European Union's Horizon 2020 research and innovation programme under grant agreement No 739500. EST is an ambitious project to build a 4-m class solar telescope, to be erected in the Canary Islands. The project is promoted by the European Association for Solar Telescopes, formed by 23 research institutions from Austria, Belgium, Croatia, Czech Republic, France, Germany, Great Britain, Greece, Hungary, Italy, Netherlands, Norway, Poland, Slovakia, Spain, Sweden, and Switzerland. Part of this work was supported by the Czech institutional project RVO:67985815. 

\section*{References}

\bibliographystyle{elsarticle-harv}
\bibliography{manuscript}

\end{document}